\documentclass[12pt]{article}
\usepackage{geometry}
\usepackage{graphicx}
\usepackage{sidecap}
\sidecaptionvpos{figure}{c}

\usepackage{wrapfig}
\usepackage[utf8]{inputenc}
\usepackage{marginnote}
\usepackage{geometry}
\usepackage{physics}
\usepackage{xcolor}
\usepackage{subcaption} 
\usepackage{amssymb}
\usepackage{yfonts}

\title{\bf On  measures of classicality/quantumness in  quasiprobability representations 
of finite-dimensional quantum systems}

\author{N.Abbasli$^1$,  
V.Abgaryan$^{1,2}$,
M.Bures$^{1,3}$,
A.Khvedelidze$^{1,4,5}$,\\
I.Rogojin$^1$
and A.Torosyan$^1$}

\date{%
 $^1$\small Laboratory of Information Technologies, JINR, Dubna, Russia\\%
 $^{2}$A.I. Alikhanyan National Science Laboratory (YerPHI), Yerevan, Armenia\\
 $^{3}$Institute of Experimental and Applied Physics, CTU, Prague, 
 Czech Republic \\
 $^4$A Razmadze Mathematical Institute, TSU, Tbilisi, Georgia \\
 $^5$Institute of Quantum Physics and Engineering Technologies, GTU, Tbilisi, Georgia \\[2ex]
}

\begin{document}

\maketitle

\begin{abstract}
  In the present report we discuss measures of classicality/quantumness of states of finite-dimensional quantum systems, which are based on a deviation of quasiprobability distributions from true statistical distributions.
Particularly, the dependence  of the global indicator of classicality  on the assigned geometry of a quantum state space is analysed for a whole family of Wigner quasiprobability representations.
General considerations are exemplified by constructing the global indicator of  classicality/quantumness for the Hilbert-Schmidt, Bures and Bogoliubov-Kubo-Mori ensembles of qubits and qutrits. 
\end{abstract}


\section{Background and motivation}

A centenary history of development of quantum theory  shows a persistent request for  a genuine unification of basic quantum-mechanical principles with concepts of classical statistical physics. The primary difficulties on this way are due to a fundamental ban originating from  the Heisenberg canonical commutation relations, $[q,p] = {\imath}\,\hbar\,,$ between the phase-space variables  $(q, p)$\,.
The  non-vanishing Plank constant $\hbar$ 
impedes the existence of a function $W_\varrho(p,q)\,,$  playing the role of  a proper  joint probability distribution of  coordinates  $q$ and momenta $p $ associated with a given quantum state $\varrho$. 
In the early years of development of quantum theory,  rejection of a complete statistical description allowed Weyl and Wigner to formulate a phase-space representation of quantum mechanics with a quasiprobability function $W_\varrho(p,q)\,,$ such that the corresponding marginals are true probability distributions of the canonically conjugate coordinates $q$ and $p $ \cite{Weyl1928,Wigner1932}.
However, in contrast to proper distributions the function $W_\varrho(p,q)$   
is not everywhere non-negative for all quantum states and thus it can only be interpreted  as a quasiprobability distribution function.\footnote{Certainly, for some states there exists such a  true statistical distribution. For example,  according  to the Hudson's  theorem  \cite{Hudson1974},  a Gaussian wave function is the only pure state corresponding to a positive Wigner function.}
Today, despite this drawback, a description of quantum systems
using the technique of quasiprobability distribution became an  important  source of our understanding of quantum phenomena (see  e.g. \cite{SperlingWalmsley2018,VeitchFerrieGrossEmerson2012,Ferrie2011} and references therein). Furthermore,  perceiving the inevitability of the existence of negative values in quasiprobability representation of states  as a reflection of  the real ``quantumness'' of  physical systems, studies move to a practical context,  where  
the negativity of states is taken as a basis for building corresponding measures of nonclassicality
(e.g., \cite{KenfackZyczkowski2004} and references therein). However, elaborating this idea we are faced with a serious complexity.
Indeed, introducing  an admissible indicator of classicality/quantumness, the following general requirements should be taken into account:
\begin{enumerate}
    \item[(I)]independence of an indicator from a representation of  quantum states;
    \item[(II)]independence of an indicator from a quasiprobability representation of states.
\end{enumerate}
While satisfying the first requirement is a relatively simple issue, it is enough to assume that an indicator is a function of a state unitary invariants, the  second task is a highly nontrivial one. There exist infinitely many quasiprobability distributions  and  a quantum state can be negative in one representation and positive in another.
In \cite{Spekkens2008}  
it was  argued  that the positivity  in one representation is neither a necessary nor a sufficient condition for classical description, nor the negativity of a specific representation is sufficient for nonclassicality.  Considering any one of these quasiprobability representations we are not able to determine absolute criteria for the classicality/quantumness. Ideally, in order to quantify a state classicality/quantumness we need to determine  characteristics which are unique for a complete family of such representations.

In the present report we will discuss both issues, (I) and (II),  constructing a 
classicality/quantumness  measure for a family of the Wigner quasiprobability representation of  finite-dimensional quantum systems. 
We will follow approach  \cite{Stratonovich,AKh2017,AKhT2018} to the construction of the Wigner quasiprobability distributions  $W^{(\boldsymbol{\nu})}_\varrho(\Omega_N)$ of an $N\--$level quantum system via the dual pairing, \begin{equation}
\label{eq:WignerFunction}
W^{(\boldsymbol{\nu})}_\varrho(\Omega_N) = \mbox{tr}\left[\varrho
\,\Delta(\Omega_N\,|\,\boldsymbol{\nu})\right]\,,
\end{equation} 
of the density matrix $\varrho \in \mathfrak{P}_N$ from the quantum state space $\mathfrak{P}_N$: 
\begin{equation}\label{eq:StateSpace}
    \mathfrak{P}_N =\{ X \in M_N(\mathbb{C}) \ |\ X=X^\dagger\,,\quad  X \geq 0\,,  \quad \mbox{tr}\left( X \right) = 1   \}\,,
\end{equation}
and the Stratonovich-Weyl (SW) kernel $\Delta(\Omega_N\,|\,\boldsymbol{\nu}) \in \mathfrak{P}^\ast_N\,$ from the dual space $\mathfrak{P}^\ast_N$:
\begin{equation}
\label{eq:SWspace}
    \mathfrak{P}^\ast_N=\{ X \in M_N(\mathbb{C}) \ |\ X=X^\dagger\,,\quad \mbox{tr}\left( X \right) = 1\,, 
    \quad
   \mbox{tr}\left( X^2 \right) = N 
    \}\,.
\end{equation}
Analysing algebraic equations (\ref{eq:SWspace}), one can conclude that,
\begin{enumerate}
    \item[a)] The phase-space $\Omega_N$\, can be  identified as a complex flag manifold,  $
\Omega_N \,\to\, \mathbb{F}^N_{d_1,d_2, \dots, d_s}={U(N)} / {H}\,, $ where 
$(d_1, d_2, \dots, d_s)$ is a sequence of positive integers with sum $N $, such that  $k_1=d_1$ and $k_{i+1}=d_{i+1}-d_i$ with $d_{s+1}=N\,.$ The corresponding  SW kernel has the  isotropy group $ H={U(k_1)\times U(k_2) \times U(k_{s+1})}\,;
$
\item[b)]The isotropy group $H$ of SW kernel provides the existence of a family of  Wigner distributions. 
The corresponding  moduli space $\mathcal{P}_N$ represents  a spherical polyhedron   on   
$(N-2)\--$dimensional sphere  $\mathbb{S}_{N-2}(1)\,$ of radius one.
Further in the text, the  $s$-dimensional moduli parameter  $\boldsymbol{\nu}=(\nu_1, \nu_2, \dots, \nu_s ) , \,s \leq N-2\, $  
will be used to enumerate the Wigner distributions (see details in  \cite{AKh2017},\cite{AKhT2018}).
\end{enumerate}
The representation independent  characteristic of  the classicality can be constructed by averaging over the moduli space  $\mathcal{P}_N(\boldsymbol{\nu})$ :
\begin{equation}
\label{eq:averageQ}
    \langle Q\rangle =
    \frac{1}{\mathrm{Vol}(\mathcal{P}_N)} \int_{\mathcal{P}_N}
    \mathrm{d}\mathcal{P}_N(\boldsymbol{\nu})\, \mathcal{Q}_N[\mathrm{g}\,|\,\boldsymbol{\nu}]
\end{equation}
of the \textit{global indicator of classicality/quantumness}  $\mathcal{Q}_N $ defined as the relative volume ratio of the subspace $\mathcal{O}[\mathfrak{P}_N^{(+)}]$ of the orbit space    $\mathcal{O}[\mathfrak{P}_N] ={\mathfrak{P}_N }/{SU(N)} \,$ of the state space  $\mathfrak{P}_N \,,$
where the Wigner function is non-negative 
\cite{AKhT2019}:
\begin{equation}
\label{eq:indQ}
\mathcal{Q}_N[\mathrm{g}\,|\,\boldsymbol{\nu}] =
\frac{ \displaystyle{\idotsint_{\mathcal{O}[\mathfrak{P}_N^{(+)}]}\,\mathrm{dP}_N(
\mathrm{g}|\boldsymbol{r})}}
{\displaystyle{\idotsint_{ \mathcal{O}[\mathfrak{P}_N]}\,\mathrm{dP}_N(
\mathrm{g}|\boldsymbol{r})}}\,.
\end{equation}
The total  orbit space $\mathcal{O}[\mathfrak{P}_N]$   
can be realised as the ordered $(N-1)$\--simplex  
in the space  of eigenvalues $\boldsymbol{r}^\downarrow =\{r_1, r_2, \dots, r_N\}$ of a density matrix $\varrho $:
\begin{equation}
\label{eq:orderedsymplex}
\mathcal{C}^{(N-1)} = \{\  \boldsymbol{r} \in \mathbb{R}^N \, \biggl| \, 
\sum_{i=1}^{N} r_i = 1, \quad 1\geq r_1\geq r_2 \geq \dots \geq r_{N-1}\geq r_N \geq 0 \ \}\,,
\end{equation}
while according to \cite{AKhT2019} the subspace  $\mathcal{O}[\mathfrak{P}_N^{(+)}]\,$   represents  a dual cone  of 
$\mathcal{O}[\mathfrak{P}_N]$:
\begin{equation}
\label{eq:OP+}
\mathcal{O}[\mathfrak{P}_N^{(+)}] =  \
\left\{\, \boldsymbol{\pi} \in \mbox{\bf spec}\left(\Delta(\Omega_N)\right) \ \, |\ \,  ( \boldsymbol{r}^\downarrow, \boldsymbol{\pi}^\uparrow) \geq 0,  \quad \forall\, \boldsymbol{r} \in \mathcal{O}[\mathfrak{P}_N]\, \right\}\,,
\end{equation}
with the cone defined via the dual pairing
\(
( \boldsymbol{r}^\downarrow, \boldsymbol{\pi}^\uparrow) =
r_1\pi_{N}+r_2\pi_{N-1} +\dots +r_{N}\pi_1\,,
\)
of $\boldsymbol{r}$ and 
the $N\--$tuple 
$\boldsymbol{\pi}$ of increasing eigenvalues of SW kernel $\Delta(\Omega_N\,|\,\boldsymbol{\nu})\,.$  

The suggested measure of classicality (\ref{eq:averageQ}) fulfils  both conditions, (I) and (II).
By the averaging procedure in (\ref{eq:averageQ}) we fulfil requirement (II) and since the global indicator of classicality/nonclassicality is defined  on the orbit  space of a quantum system and therefore provides an unitary invariant 
measure, the requirement (I) is satisfied as well. However,    the indicator (\ref{eq:averageQ})  depends on metrical characteristics of the 
moduli space $\mathcal{P}_N$ and the orbit space $\mathcal{O}[\mathfrak{P}_N]\,.$
Since the moduli space is represented by an $(N-2)\--$dimensional 
spherical polyhedron, we suppose that the corresponding measure corresponds to uniform distribution on $\mathbb{S}_{N-1}(1)$, while 
the measure on the orbit space, $\mathrm{dP}_N(
\mathrm{g}|\boldsymbol{r})=\sqrt{\det||\mathrm{g}||}\, \mathrm{d}r_1\wedge \mathrm{d}r_2\wedge\cdots \wedge\mathrm{d}r_{N}\,,$ is induced from the  Riemannian metric $\mathrm{g}$ on the state space $\mathfrak{P}_N \,.$
In the remaining part of the report we will discuss the  dependence of the suggested indicator of classicality on the metric  of a quantum state space. 
From  a wide variety of special Riemannian metrics commonly used in the Quantum Statistics and Information Theory,  we will analyse the Hilbert-Schmidt metric and two monotone metrics \cite{MorozovaChensov1990,PetzSudar1996}), the Bures and the Bogoliubov-Kubo-Mori
metrics. Detailed discussion of   the indicator of classicality for  qubit  $(N=2)$ and qutrit $(N=3)$ will be given.

\section{Riemannian geometry of state space  $\mathfrak{P}_N$ and    classicality/quantumness  indicator  $\mathcal{Q}_N$}

In this section the results of our studies of the dependence of the indicator (\ref{eq:indQ}) on the metric of a quantum state space will be presented. We will consider a basic,  Hilbert-Schmidt (HS) metric and two representatives of the family of the so-called monotone metrics, the Bures (B) and  Bogoliubov-Kubo-Mori (BKM) metrics.

\noindent$\bullet$ {\bf The Hilbert-Schmidt metric} $\bullet$\,
For an $N\--$dimensional quantum system
the infinitesimal version  of the Hilbert-Schmidt distance is given by the expression:  
\begin{equation}\label{eq:HSmetric}
\mathrm{g_{\mathrm{HS}}} = 4\, \mathrm{tr}\left(\mathrm{d}
\varrho\otimes\mathrm{d}\varrho
 \right)\,.
\end{equation}
For further computational aims it is convenient to rewrite  (\ref{eq:HSmetric}) in terms of SVD of a density matrix,  
\(
  \varrho = UDU^{\dag}\,,  
\)
where $U \in SU(N)\,$, and $D=\mbox{diag}||r_1, r_2, \dots, r_N||$\, with descending order of eigenvalues from the  $(N-1)$\--simplex  (\ref{eq:orderedsymplex}).
Here we assume that the spectrum of  $\varrho$
is generic and thus the arbitrariness of $U$ is given by the isotropy group  represented by the torus $T$ of $SU(N)$. In terms of SVD coordinates, 
the volume form  factorizes
into the measure $\omega_{{}_\mathrm{SU(N)/T}}$ on the coset ${SU(N)/T}$ induced from the Haar measure on the $U(N)$ group manifold and the 
  ``radial part'' factor which depends on the spectrum of a state only. The latter represents  the Hilbert-Schmidt measure  $\mathrm{dP}(\mathrm{g}_{{}_\mathrm{HS}})$ on the orbit space (\ref{eq:orderedsymplex}), 
\begin{equation}
\label{eq:HSmeasureOS}
    \mathrm{dP}_N(\mathrm{g}_{{}_\mathrm{HS}}|\boldsymbol{r})= c_{{}_\mathrm{HS}}\,\delta(\sum_{i=1}^N \,r_i -1)\, \prod_{i<j}^N\,{(r_i-r_j)^2}\,\mathrm{d}r_1\wedge \mathrm{d}r_2\wedge\cdots \wedge\mathrm{d}r_{N}\,,
\end{equation}
where $c_{{}_\mathrm{HS}}$ is a normalization constant.

\noindent$\bullet$ {\bf The Bures and  Bogoliubov-Kubo-Mori metrics} $\bullet$\,
Using the SVD decomposition of elements of $\mathfrak{P}_N$\,,  the stochastically monotone  metrics can be written in the following form:
\begin{equation}
 \label{eq:MonotoneM} 
    \mathrm{g}_f= \frac{1}{4}\sum_{i=1}^N\,
    \frac{\mathrm{d}r_i
    \otimes\mathrm{d}r_i}{r_i}\,  + \frac{1}{2}\,
    \sum_{i< j}^N\,c_f(r_i\,, r_j){(r_i-r_j)^2}\,
    \left(U^\dagger\mathrm{d}U\right)_{ij}\otimes
  \left(U^\dagger\mathrm{d}U\right)_{ij} \,,
\end{equation} 
where $c(x,y)$ is the so-called Morozova-Chentsov function; $c_f(x,y) = \frac{1}{yf(x/y)}$ is given by the operator monotone function $f(t)$. For the Bures and BKM metrics these functions are $f_{\mathrm{BW}}(t) =(1+t)/{2} $ and $ f_{{}_\mathrm{BKM}}(t)= {(t-1)}/{\ln{t}}\,$, respectively.
Having these representations, we are in a position to compare the indicators $\mathcal{Q}$ for the simplest, two- and three-level systems endowed with the above described metrics.    

\subsection{Qubit}

From (\ref{eq:SWspace}) it follows that the spectrum of SW kernel of a 2-level system is unique, 
$
\mbox{\bf spec}\left(\Delta_2\right) =\{ (1+\sqrt{3})/{2}\,, (1-\sqrt{3})/{2}\}\,.
$
Its dual pairing with  a  2-level density  matrix  $\rho=\frac{1}{2}\left[\,
\mathbb{I}_2+
(\boldsymbol \xi\,, \boldsymbol{\sigma})\right]\,,$ characterized by the Bloch vector $\boldsymbol{\xi}\ \in\mathbb{R}^3\,,$ 
gives the Wigner quasiprobability distribution of the qubit defined on a 2-sphere: 
 $$
W_\varrho(\boldsymbol{n})=\frac{1}{2} + \frac{\sqrt{3}}{2}\,
     (\boldsymbol{\xi},\boldsymbol{n})\,, \qquad 
     \boldsymbol{n} \in \mathbb{S}^2\,.
$$
All mixed states belong to the Bloch ball, 
$
 (\boldsymbol{\xi}\,, \boldsymbol{\xi}) \leq 1\,,$
while the positivity cone (\ref{eq:OP+}) represents  qubit states  inside the following ball:  \(
   (\boldsymbol{\xi}\,, \boldsymbol{\xi}) < 1/3\,.
\)

\noindent$\bullet$ {\bf The Hilbert-Schmidt metric} $\bullet$\,
Taking into account the positivity domain (\ref{eq:OP+}) and  using the expression (\ref{eq:HSmeasureOS}) for $N=2$, the indicator  $\mathcal{Q}_{2}$ of the Hilbert-Schmidt qubit reduces to the ratio of two simple integrals,  
\begin{equation}
    \mathcal{Q}_2 [\mathrm{g_{{}_{\mathrm{HS}}}}] 
    \,=\,
    \frac{
    {\int_{0} ^{\frac{1}{\sqrt{3}}} r^2 dr}}{
    {
    \int_{0}^{1} r^2 dr}} = \frac{1}{3\sqrt{3}}\approx
    0.19245\,.
\end{equation}

\begin{SCfigure}
 \caption{A qubit probability  $Q(r)\,,$  calculated for the Hilbert-Schmidt, Bures  and the  Bogoliubov-Kubo-Mori metrics.}
 \includegraphics[width=0.64\textwidth]{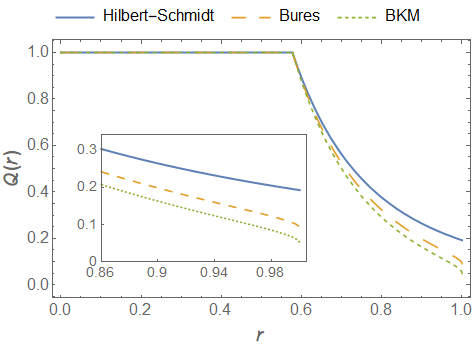}
\label{Fig:QAll}
\end{SCfigure}

\noindent$\bullet$ {\bf Bures and BKM metric} $\bullet$\,
Similar calculation for the  Bures and BKM ensemble of qubits give, 
\begin{eqnarray}
\label{eq:QubitB+BKM}
     \mathcal{Q}_2 [\mathrm{g_{{}_{\mathrm{B}}}}]  \,&=&\,
    \frac{\mathrm{Vol_B}(\frac{1}{\sqrt{3}})}{\mathrm{Vol_B}(1)} =
    \frac{2}{\pi}\left[\arcsin\frac{1}{\sqrt{3}} - \frac{\sqrt{2}}{3}\right]\approx 0.09172 \,,\\
  \mathcal{Q}_2 [\mathrm{g_{{}_{\mathrm{BKM}}}}]  \,&=&\frac{\mathrm{Vol_{{}_{BKM}}}(\frac{1}{\sqrt{3}})}
  {\mathrm{Vol_{{}_{BKM}}}(1)} =\frac{2}{\pi}\left[ \arcsin\frac{1}{\sqrt{3}}-\sqrt{\frac{2}{3}} \operatorname{arcoth}\sqrt{3}\right]\approx 0.0495506\,, \nonumber 
\end{eqnarray}
where $\mathrm{Vol_X}(r)$ denotes the volume of the Bloch ball of radius $r$ in metric $``X"\,.$
The corresponding probability $Q(r)\,$  to find a qubit state $\varrho$  with positive WF within the Bloch ball of radius  $r\,$ is depicted in  Fig.\ref{Fig:QAll}.

\subsection{Qutrit}

\noindent$\bullet$ {\bf The Hilbert-Schmidt metric} $\bullet$\, According to (\ref{eq:SWspace}), the Wigner quasiprobability representation  of a  3-level system is one-parametric. The spectrum of SW kernel can be parametrized by the apex angle $\zeta  \in [0, \pi/3]$ of a unit circle segment~\cite{AKh2017}:  
\begin{eqnarray}
\label{eq:specDelta}
\mbox{\bf spec}\left(\Delta_3\right) =\biggl\{
\frac{1}{3}+\frac{2}{\sqrt{3}}\sin\zeta+\frac{2}{3}\cos\zeta,\,
\frac{1}{3}-\frac{2}{\sqrt{3}}\sin\zeta+\frac{2}{3}\cos\zeta,\, 
\frac{1}{3}-\frac{4}{3}\cos\zeta\,
\biggl\}.
\end{eqnarray}
Decomposing a qutrit density matrix spectrum via the polar coordinates  $(r, \varphi)$,
\begin{eqnarray}
\label{eq:specrho}
\mbox{\bf spec}\left(\varrho\right) =\biggl\{
\frac{1}{3}-\frac{2r}{\sqrt{3}}
\cos\frac{\varphi+2\pi}{3},
\frac{1}{3}-\frac{2r}{\sqrt{3}}
\cos\frac{\varphi+
4\pi}{3},
\frac{1}{3}-\frac{2r}{\sqrt{3}}
\cos\frac{\varphi}{3} 
\biggl\},
\end{eqnarray}
a qutrit  orbit space and its subspace of WF positivity reads
($r \geq 0\,,  \varphi \in [0, \pi]$):
\begin{eqnarray}
 \label{eq:OrbitQutrit}
 \mathcal{O}[\mathfrak{P}_3]\,:  \cos\left(\frac{\varphi}{3}\right) \leq \frac{1}{2\sqrt{3}r}\,, 
 \quad 
\mathcal{O}[\mathfrak{P}^{(+)}_3]\, :  \ \cos\left(\frac{\varphi}{3} +\zeta -\frac{\pi}{3}\right) \leq \frac{1}{4\sqrt{3}r}\,.
 \label{eq:PosiDomainrphi}
 \end{eqnarray}
Taking into account the expression for the Hilbert-Schmidt measure  on the orbit space $\mathcal{O}[\mathfrak{P}_3]$,  
\(
\omega_{\mathcal{O}[\mathfrak{P}_3]}  =
 r^7 \sin^2{\varphi}\,
\mathrm{d}r\wedge\mathrm{d}\varphi\,,
\)
we derive the global indicator of classicality of the Hilbert-Schmidt qutrit as function of the moduli parameter $\zeta\,$\cite{AKhT2019}:
\begin{eqnarray}
\label{eq:Ratio}
\mathcal{Q}_3 (\zeta)= \displaystyle{
\frac{\int_0^\pi d\varphi\int_{0} ^{\frac{1}{4\sqrt{3} \cos{\left(\frac{\varphi}{3}+\zeta-\frac{\pi}{3}\right)}}} r^7 \sin^2(\varphi) dr}{
\int_0^\pi d\varphi\int_{0} ^{\frac{1}{2\sqrt{3} \cos{\frac{\varphi}{3}}}} r^7 \sin^2(\varphi) dr }}=\frac{1}{128}\,\frac{1+20 \cos^2{\left(\zeta -{\pi }/{6}\right)}}{ \left(-1+4\cos^2{\left(\zeta -{\pi }/{6}\right)}\right)^5}\,.
\end{eqnarray}

Note, that the indicator $\mathcal{Q}_3(\zeta)$ attains at a qutrit moduli parameter 
$\zeta = {\pi}/{6}$ the absolute minimum, $\min_{\zeta \in [0,\frac{\pi}{3}]}\mathcal{Q}_3 (\zeta)\approx 0.000675\,,$
corresponding to SW kernel with the spectrum:  
\(
\mbox{\bf spec}\left(\Delta_3\right)|_{\zeta=\frac{\pi}{6}}=||\,\frac{1+2\sqrt{3}}{3}\,, \frac{1}{3}\,, \frac{1-2\sqrt{3}}{3}\,||\,.
\)

\section{Final remarks}

As it was outlined in the first part of our report, a true classicality/quantumness measure, being universal for different quasiprobability representations, may be sensitive to the geometry of a state space. 
Our calculations of the average of the global indicator $Q(\zeta)$ over a qutrit moduli space   
support this  supposition,
\begin{equation}
 \langle Q_{{}_{\mathrm{HS}}}\rangle_\zeta = 0.00136368\,, \qquad
 \langle Q_{{}_{\mathrm{B}}}\rangle_\zeta =0.00019165\,,  \qquad
 \langle Q_{{}_{\mathrm{BKM}}}\rangle_\zeta=0.00002762\,.
\end{equation}

\noindent {\bf Acknowledgment} The work of MB was supported in part by  the EU Regional Development Fund-Project No. CZ.02.1.01/0.0/0.0/16\_019/0000766.


\end{document}